%% file: ms_rev.tex
\documentclass[aps,twocolumn,pra,superscriptaddress,nofootinbib,amsmath,amssymb,floats,floatfix,english,longbibliography]{revtex4-2}
\usepackage{dsfont}
\usepackage{graphicx}
\usepackage{dcolumn}
\usepackage{bm}
\usepackage{latexsym,epsfig}
\usepackage{graphicx}
\usepackage{verbatim}
\usepackage{comment}
\usepackage{amsmath}
\usepackage{amssymb}
\usepackage{stmaryrd}
\usepackage{color}

\usepackage[colorlinks=true,linkcolor=blue,citecolor=blue]{hyperref}

\begin{document}
\title{Pairs, trimers and   BCS-BEC crossover near a flat band: \\
the sawtooth lattice}
\author{Giuliano Orso}
\email{giuliano.orso@u-paris.fr}
\affiliation{Universit\'{e} Paris Cit\' e, Laboratoire Mat\'{e}riaux et Ph\'{e}nom\`{e}nes Quantiques (MPQ), CNRS, F-75013, Paris, France}
\author{Manpreet Singh}
\affiliation{Centre for Quantum Engineering Research and Education,
	TCG Centres for Research and Education in Science and Technology,
	Sector V, Salt Lake, Kolkata 700091, India}
\affiliation{Universit\'{e} Paris Cit\' e, Laboratoire Mat\'{e}riaux et Ph\'{e}nom\`{e}nes Quantiques (MPQ), CNRS, F-75013, Paris, France}

\date{\today}

\begin{abstract}
We investigate pairing and superconductivity in the attractive Fermi Hubbard model on the one-dimensional sawtooth lattice, which exhibits a flat band by fine-tuning the hopping rates. We first  solve the
two-body problem, both analytically and numerically, to extract the binding energy and the effective mass of the pairs. Based on the DMRG method, 
we address the ground-state properties of the many-body system, assuming equal spin populations. We compare  our results with those available for a linear chain, where the model is  integrable by Bethe ansatz, and show that the multiband nature of the system  substantially modifies the physics of the BCS-BEC crossover.
Near a flat band, the chemical potential  remains always close to its zero-density limit predicted by the two-body physics. In contrast, the  pairing gap exhibits a remarkably strong density dependence and, differently from the pair binding energy, it is no longer peaked at the flat-band  point. 
We show that these results can be interpreted in terms of polarization screening effects, due to an anomalous
 attraction between pairs in the medium and single fermions.  Importantly,  we unveil that  three-body bound states (trimers) exist in the sawtooth lattice, in sharp contrast with the linear chain geometry, and we compute their binding energy. The nature of these states is investigated via a strong coupling variational approach, revealing that they originate from tunneling-induced exchange processes.
\end{abstract}

\maketitle

\section{Introduction}
During the last ten years there has been a growing interest 
on FB lattices \cite{Leykam:2018}. These are periodic systems, described by tigth binding models, in which one or more dispersion relations is flat or almost flat.  The corresponding eigenstates are localized on few lattice sites due to destructive quantum interference.
The absence of kinetic energy together with the inherent
macroscopic degeneracy make flat-band (FB) systems ideal candidates to enhance interaction effects. 
For instance they  provide a viable route to enhance the 
superconducting transition temperature~\cite{Khodel:JETP1990,Kopnin:PRB2011,Heikkila2011,Aoki:2020}, generate fractional quantum Hall states at room temperature \cite{Tang:PRL2011}, and produce many other intriguing quantum effects.  

Lattice models containing a flat band have been realized experimentally with optical
lattices for ultracold atoms \cite{JO:PRL2012,Taie:ScAdv2015,Leung:PRL2020}, 
photonic lattices \cite{Gersen:PRL2005,Mukherjee:PRL2015},
semiconductor microcavities \cite{Jacqmin:PRL2014} 
and artificial electronic lattices \cite{Drost2017,Slot2017,Huda:PRR2020}.
The recent discovery \cite{Cao2018} of unconventional superconductivity and strongly correlated phases in bilayer graphene twisted at a magic angle, causing the emergence of flat bands in 
the electronic structure, has further boosted the theoretical and experimental research on FB sistems.

The physics of two-body bound states in the presence of a flat band has been recently explored theoretically in different contexts, including topological matter  \cite{Salerno:PRR2020,Kuno:PRA2020,Flannigan_2020,Pelegri:PRR2020} and the link between the inverse effective mass of the bound state and the quantum metric of the single-particle states  \cite{Torma:PRB2018,Iskin:PRA2021,Iskin:arXiv2021}.  
This second direction is related to the more general question of understanding how transport and superconductivity can occur in system with quenched kinetic energy \cite{Peotta2015,Julku:PRL2016,Mondaini:PRB2018,Iskin:PRA2019b,Balents2020,Vermae:PNAS2021,chan2021pairing,Pykkonen:PRB2021,Hofmann:PRB2020,Peri:PRL2021}. 
By increasing the fermion-fermion attraction, the many-body system progressively transforms into a bosonic gas of diatomic molecules. The evolution from a  Bardeen–Cooper–Schrieffer (BCS) state to a Bose-Einstein Condensate (BEC), commonly referred to as the BCS-BEC crossover, has been investigated both theoretically and experimentally in single-band dispersive systems, going from superconductors  \cite{Micnas:RMP1990}  to atomic Fermi gases  \cite{Bloch:RMP2008}. Recent theoretical works   
\cite{Iskin:PRA2016,He:PRA2016,Xu:PRA2016} have generalized the theory to two-band \emph{continuous} models  
describing superfluid  Fermi gases near an orbital Feshbach resonance, and a significant increase of the critical temperature $T_c$ has been predicted when the  lower band becomes shallow \cite{Tajima:PRB2019,Tajima:condmat2020}. 
Transport in many-body bosonic flat-band systems has also been explored, see for instance \cite{Huber:PRB2010,You:PRL2012,Phillips:PRB2015,BabouxPRL2016,Ozawa:PRL2017}.

\begin{figure}
\includegraphics[clip,width=0.95\columnwidth]{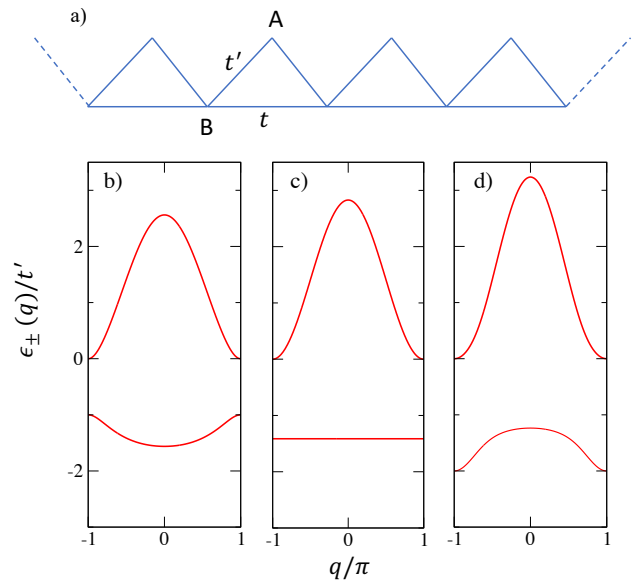}
\caption{(Color online) (a) Geometry of the 1D sawtooth lattice. The lattice containing two sites per unit cell, indicated by $A$ and $B$. We indicate by $t^\prime$ the hopping rate between A and B sites and by $t$ the hopping rate between $B$ sites.  Notice that there is no tunneling between A sites.
	Panels (b-d): Dispersion relations of the two bands for different values of the tunneling ratio: $t/t^\prime =1/2$ (b), $t/t^\prime =1/\sqrt 2$ (c), corresponding to the flat-band point,  and $t/t^\prime=1$ (d).
 In this work we fix the energy scale by setting $t^\prime=1$. 
}
\label{fig:fig1_lattice}
\end{figure}

 In this work we study,  in a unified framework, the influence of the multiband structure and the vicinity to the flat band on pairing phenomena, going from the formation of molecules in vacuum to  superconductivity in many-body fermionic systems.  Our investigation is based on the attractive Fermi Hubbard model on the
 one-dimensional (1D) sawtooth lattice (also known as triangular or Tasaki lattice),
  shown in Fig.~\ref{fig:fig1_lattice} (a).
Its unit cell contains two lattice sites, called A and B; particles can hop between two B sites with  rate $t$, while tunneling between A and B sites occurs at a rate $t^\prime$. 
Two combinations of the tunneling rates are of special interest: \emph{i)} the FB point, corresponding to $t=t^\prime/\sqrt 2$, where the
	lower Bloch band becomes dispersionless, and  \emph{ii)} $t=0$, where the sawtooth lattice 
	reduces to the linear chain and the Hubbard model  is then integrable by Bethe ansatz.  
We first present a thorough solution of the two-body problem, from which we extract the binding energy and the effective mass of the pair as a function of the  tunneling rates $t, t^\prime$ and the Hubbard strength $U$.   
We then show that in many-body systems the proximity to a flat band strongly modifies the nature of the superconducting state, as compared to the integrable limit. The chemical potential remains always closed to its zero-density limit, even in the weakly interacting regime. In contrast,   the superfluid pairing gap is strongly depleted at finite density and its peak is shifted with respect to the FB point. We explain this surprising effect by studying  the change in the ground state energy of the system upon adding an extra fermion. For nonzero $t$ and $|U|$ sufficiently large,  this quantity falls below the bottom of the single-particle energy spectrum,
indicating that  pairs and single fermions tend to attract each other. 
To support this picture, we explicitly show that three-body bound states  do appear in the sawtooth lattice. We compute their binding energy $E_b^\textrm{trim}$ as a function of the interaction strength and the tunneling rates, and show that  $E_b^\textrm{trim}$  exhibits a peak at the FB point, in complete analogy with the two-body case. Importantly, we use a strong coupling variational approach to show that trimers originate from tunneling-induced exchange processes. 

Superconductivity in the sawtooth lattice at the FB point has been recently investigated numerically in 
Ref.\cite{chan2021pairing}, with a focus on  the superfluid  weight $D_s$. The authors  introduced a modified multiband BCS theory with sublattice-dependent order parameters to 
account for the different connectivity of the A and B sites. In this way the mean field approach was shown to compare well with   density matrix renormalization group (DMRG) calculations.  

The article is organized as follows. In Sec.~\ref{sec:formalism} we review the single-particle properties of the sawtooth lattice and present the formalism used to solve the two-body problem in a  multiband lattice. In Sec.~\ref{sec:boundstates} we show our results for the binding and the effective mass of the two-body bound states, both at the FB point and for generic tunneling rates.
 In Sec.\ref{sec:manybody} we present our DMRG results for the BCS-BEC crossover at finite density, 
 while in Sec.~\ref{sec:3body} we discuss the formation of trimers  in the sawtooth lattice. Finally in Sec.\ref{sec:conclusions}  we present our conclusions. 

 \section{Theoretical approach}
 \label{sec:formalism}

\subsection{Single-particle properties}
We recall here the single-particle properties of the 1D sawtooth lattice, shown in Fig.~\ref{fig:fig1_lattice} (a). 
The  tight-binding Hamiltonian is given by  
\begin{equation}\label{Hsp}
H_\textrm{sp}=\sum_i t |i^B \rangle \langle i+1^B|+ \textrm{h.c.} + t^\prime |i^A\rangle ( \langle i^B| +\langle i+1^B|)+\textrm{h.c.},
\end{equation}
with $|i^{A/B} \rangle$ denoting the local (site) basis.
The  dispersion relations of the two bands associated to the Hamiltonian (\ref{Hsp}) are given by
\begin{equation}\label{disp}
\varepsilon_{\pm}(q)=t \cos q \pm \sqrt{t^2 \cos^2 q+2t^{\prime 2} (1+\cos q)},
\end{equation}
where $q$ is the wave-vector of the Bloch state and we have set to one the distance between two adjacents  sites.  
In Fig.~\ref{fig:fig1_lattice} (panels b-d) we show how the shapes of the two bands evolve as the
tunneling ratio  $t/t^\prime$ is changed. While the upper band  is always concave down at $q=0$,
 the bottom $q=q_\textrm B$ of the lower band changes from $q_\textrm B=0$, for $t/t^\prime<1/\sqrt{2}$, to  $q_\textrm B=\pi$ for $t/t^\prime >1/\sqrt{2}$. 
Exactly at  $t/t^\prime =1/\sqrt{2}$, the lower band becomes flat, $\varepsilon_{-}(q)=-\sqrt 2 t^\prime $ (c), implying that  the  inverse effective mass  $1/m^*=\epsilon_{-}^{\prime \prime}(q_\textrm{B})$  vanishes. For any other value of $t$ Eq.~(\ref{disp}) yields 
\begin{equation}\label{eq:m*}
	\frac{1}{m^*}=\begin{cases}
		-t+\frac{t^{\prime 2}+t^2}{\sqrt{4t^{\prime 2}+t^2}}  &  \text{if $\frac{t}{t^\prime}<\frac{1}{\sqrt{2}}$} \\
	-\frac{ t^{\prime 2}}{t}+	2t &  \text{if $\frac{t}{t^\prime}>\frac{1}{\sqrt{2}}$} .
	\end{cases}
\end{equation}

The amplitudes of the Bloch states at site $j$ associated to the energy bands $\varepsilon_{\nu}(q)$ can be conveniently written  as $\Psi_{q\nu}(j)=
 \frac{e^{iq j}}{\sqrt L} 
 \begin{pmatrix}
  \alpha_{q\nu} \\ 
  \beta_{q\nu} 
\end{pmatrix}$, where $L$ is the number of unit cells, while $\alpha_{q\nu}, \beta_{q\nu}$  satisfy  
\begin{equation}\label{fun}
 \alpha_{q\nu}=\frac{t^\prime (1+e^{-i q})}{\varepsilon_{\nu}(q)} \beta_{q\nu}, 
\end{equation}
together with the normalization condition $|\alpha_{q\nu}|^2+|\beta_{q\nu}|^2=1$.  Since $\epsilon_{\nu}(-q)=\varepsilon_{\nu}(q)$, we are free to choose $\beta_{q\nu}$ real
and satisfying $\beta_{-q\nu}=\beta_{q\nu}$. From Eq.~(\ref{fun}) we then find that $ \alpha_{-q\nu}= \alpha_{q\nu}^*$, where the star indicates the complex conjugate.

\subsection{Two-body problem}
We now consider two particles hopping on the sawtooth lattice and coupled by contact interactions.
The two-body Hamiltonian is given by $\hat H=\hat H_0+\hat U$, where  $\hat H_0=\hat H_\textrm{sp} \otimes  \mathds{1} +\mathds{1}  \otimes  \hat H_\textrm{sp}$ is the noninteracting Hamiltonian and  $\hat U=U (\hat P^A +\hat P^B)$  accounts for contact interactions between the two particles. Here
  $\hat P^\sigma= \sum_{m^\sigma} | m^\sigma m^\sigma\rangle \langle m^\sigma m^\sigma|$ are the pair projector operators 
over the doubly occupied sites of the $\sigma=A,B$ sublattices. 
The properties of two-body bound states  can be obtained by mapping the stationary Schrodinger equation into an effective single-particle model for the center-of-mass motion of the pair, as done  in Ref.s \cite{Wouters:PRA2006,Dufour:PRL2012}  for continuous and lattice models, respectively.  
If the external potential is periodic, the momentum $Q$ of the pair is conserved and the problem further reduces to finding the eigenvalues of a $N_b \times N_b$ matrix, where $N_b$ is the number of basis sites per unit cell, as we shall see below for $N_b=2$. The same equation has been recently obtained 
by Iskin in Ref.~\cite{Iskin:PRA2021} by using a different (variational) approach.  Scattering states in flat bands have instead been discussed in \cite{Valiente:JPB2017}, but can also be obtained by adapting the formalism below, as done in Ref.\cite{Orso:PRL2005}.

We start by writing  the two-body Schrodinger equation as $(E -\hat H_0)|\psi\rangle=\hat U|\psi\rangle$, where $E$ is the total energy of the pair.  Substituting it into the Schrodinger equation and bringing the operator $(E -\hat H_0)$ on the rhs yields
\begin{equation}
\label{formalism2}
\frac{1}{U}|\psi\rangle=(E -\hat H_0)^{-1} \hat P^A |\psi\rangle + (E -\hat H_0)^{-1} \hat P^B |\psi\rangle.
\end{equation}
Next, by projecting the wave-function (\ref{formalism2}) on the doubly occupied states $| m^\sigma m^\sigma\rangle$, we obtain a close equation for 
the corresponding amplitudes $f(n)=\begin{pmatrix}
  \langle n^A n^A| \psi \rangle\\ 
   \langle n^B n^B |  \psi \rangle
\end{pmatrix}$ as
\begin{equation} \label{mio}
f(n)\frac{1}{U}= \sum_m K(n,m) f(m),
\end{equation}
where, for given values of  $n$ and $m$, $K(n,m)$ is a $2\times 2$ matrix depending parametrically on the energy and whose entries are given by  $K^{\sigma\sigma^\prime}(n,m)=\langle n^\sigma n^\sigma| (E -\hat H_0)^{-1} | m^{\sigma^\prime} m^{\sigma^\prime}\rangle $. The latter can be conveniently expressed in terms of the components of the single-particle Bloch wave-functions $\Psi_{q\lambda}(j)$, so that Eq.(\ref{mio}) takes the form
\begin{equation}\label{eq:2body1}
f(n) \frac{1}{U}=\frac{1}{L^2} \sum_{m,q,p,\nu,\nu^\prime} \frac{e^{i (q+p)(n-m)}}{E-\varepsilon_\nu(q)-\varepsilon_{\nu^\prime}(p)} M_{\nu \nu^\prime}(q,p) f(m),
\end{equation}
where  
\begin{equation}
M_{\nu \nu^\prime}(q,p)=\begin{pmatrix}
  \alpha_{q\nu}\alpha^*_{q\nu} \alpha_{p\nu^\prime}  \alpha^*_{p\nu^\prime} &  \alpha_{q\nu}\beta^*_{q\nu} \alpha_{p\nu^\prime}  \beta^*_{p\nu^\prime} \\ 
\beta_{q\nu}\alpha^*_{q\nu} \beta_{p\nu^\prime}  \alpha^*_{p\nu^\prime} &  \beta_{q\nu}\beta^*_{q\nu} \beta_{p\nu^\prime}  \beta^*_{p\nu^\prime} 
\end{pmatrix}.  
\label{eq:2body2}
\end{equation}
One can easily see that the eigenstates of Eq.(\ref{eq:2body1}) are plane waves 
$f(n)=\frac{e^{i Q n}}{\sqrt{L}}f_Q$, with $Q$ being the center-of-mass momentum of the pair. By substituting it into Eq.(\ref{eq:2body1}) and taking the continuum limit, we end up with
the eigenvalue problem 
\begin{equation}\label{2x2}
f_Q \frac{1}{U} =R f_Q,   
\end{equation}
where $R=R(E,Q)$ is a $2\times 2$ matrix defined as
\begin{equation}\label{eq:integ}
R=  \sum_{\nu,\nu^\prime}\int_{-\pi}^{\pi} \frac{dq}{2\pi} \frac{ M_{\nu \nu^\prime}(q,Q-q)}{E-\varepsilon_\nu(q)-\varepsilon_{\nu^\prime}(Q-q)}. 
\end{equation}
The two eigenvalues of the 
matrix $R$ are given by
\begin{equation}\label{det}
\lambda_{\pm} =\frac{R_{11}+R_{22}}{2}\pm \frac{1}{2}\sqrt{(R_{11}-R_{22})^2+4 |R_{12}|^2}.
\end{equation}
For a given interaction strength $U$ and quasi-momentum $Q$, the energy levels of bound states 
are obtained by looking for solution of $\lambda_{\pm}(E,Q)=1/U$, the energy 
$E$ taking values outside the noninteracting two-body energy spectrum. 
In the following we fix the energy scale by setting $t^\prime=1$ and  restrict to attractively bound states, corresponding to $U<0$. 

\section{Two-body results}
\label{sec:boundstates}

\subsection{Bound states at FB point}
We present our results for the two-body bound states for the special case $t=1/\sqrt{2}$, where the lower Bloch band becomes flat, see Fig.~\ref{fig:fig1_lattice} (c).
We will be interested in the solutions of Eq.(\ref{2x2}) with energy $E<E_\textrm{ref}$, where  
$E_\textrm{ref}=2\epsilon_{-}(q_\textrm{B})=-2\sqrt 2$ is the ground state energy of the two-body 
system in the absence of interactions.
These states are often referred to as \emph{doublons}, since for large $|U|$ the two particles sit at the same site and form a tightly bound molecule with energy  $E\sim U$.

The integration over momentum in Eq.(\ref{eq:integ}) will be generally performed numerically.
Analytical integration is also possible via residue techniques, although the calculation can become  difficult for arbitrary combinations of the  parameters $E$ and $Q$. 
As an example, we provide here the exact expression for the matrix $R$ valid for zero center-of-mass momentum and  $E<E_\textrm{ref}$. This allows us to extract the pair binding energy exactly  for any $U<0$. 
To this end, we substitute  in Eq.(\ref{eq:2body2}) the amplitudes of the Bloch wavefunctions
obtained from Eq.(\ref{fun}):
\begin{eqnarray}\label{FBcomponents}
\alpha_{q-}&=&-\frac{1+e^{-iq}}{\sqrt{ 2(2+\cos q )}},\;\;
\beta_{q-}=\frac{1}{\sqrt{2+\cos q}} \\
\alpha_{q+}&=&\frac{1+e^{-iq}}{\sqrt{ 2(2+\cos q )(1+\cos q)}},\;\;
\beta_{q+}=\sqrt{\frac{1+\cos q}{2+\cos q}}, \nonumber
\end{eqnarray} 
and the corresponding dispersion relations of the two bands, with $\varepsilon_+(q)=\sqrt 2 (1+\cos q)$.
 For $Q=0$ the integration over momentum is performed  by introducing the
complex variable $z=e^{iq}$, so that the integrating function takes the form of a ratio $t(z)/y(z)$ of  two analytical functions. We then calculate the integral via the  
 Cauchy's residue theorem of complex analysis, after identifying the poles inside the circle $|z|=1$. 
 This gives   
\begin{widetext}
\begin{eqnarray}
R_{11}&=&\frac{8 \sqrt{E(E-4\sqrt 2)}+E(16 + 2 \sqrt 2 E -E^2+2 \sqrt 2 \sqrt{E^2-2}) -\frac{12E(E+\sqrt 2 )}{\sqrt{E^2-2}} } {E (32 \sqrt 2  + 24 E - E^3)} \nonumber \\
R_{12}&=&R_{21}=\frac{2 \left(\sqrt{2} \sqrt{-\frac{E \left(E^2-2\right)}{4 \sqrt{2}-E}}+\sqrt{2} E+2\right)}{\left(E+2 \sqrt{2}\right)^2 \sqrt{E^2-2}} \label{exact} \\
R_{22}&=&\frac{2  \left(\sqrt{2} E +2\right) \sqrt{-E-\sqrt{2}}+ E \left(E +\sqrt{2}\right) \sqrt{\frac{ E \left(2-\sqrt{2} E \right)}{\sqrt{2} E-8}}}{\left(E+2 \sqrt{2}\right) \sqrt{\sqrt{2} -  E} \left(E^2+3 \sqrt{2} E+4\right)}. \nonumber
\end{eqnarray}
\end{widetext}
We substitute Eq.~(\ref{exact}) in  Eq.~(\ref{det}) and obtain the exact energy of the bound state from the implicit condition  $U=1/\lambda_{-}(E)$ (notice that the eigenvalue $\lambda_{+}$ yields the energy of the first excited bound state). 
\begin{figure}
	\includegraphics[width=0.98\columnwidth]{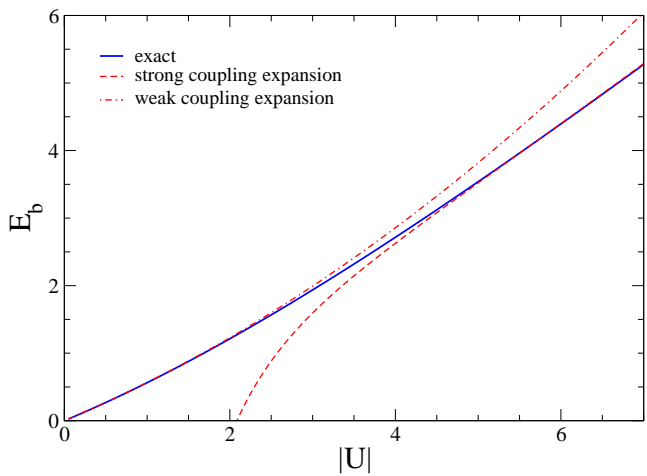}
	\caption{(Color online) Binding energy of the deepest two-body bound state at the flat band point, $t=1/\sqrt 2$, plotted as a function of the modulus of the interaction strength (blue solid line). The dot-dashed line corresponds to the weak coupling expansion, Eq.~(\ref{eq:FBweakU}), while 
		the dashed line represents the strong coupling expansion, Eq.~(\ref{EbstrongUFB}). 
		We set 
		$t^\prime=1$ as energy unit.}
\label{fig:fig2_EbFBvsU}
\end{figure}
We then extract the binding energy $E_b$  from the relation $E=-E_b+E_\textrm{ref}$, so that $E_b>0$ if the  state is bound. 
In  Fig.~\ref{fig:fig2_EbFBvsU} we plot our results for the binding energy as a function of the interaction strength  (blue solid curve).  We can use the exact relation between $E$ and $U$ to obtain the 
asymptotic expansions for the binding energy in the weak and in the strong coupling regimes.
In the noninteracting limit the binding energy vanishes, implying that $E=E_\textrm{ref}=-2 \sqrt 2$. For small $U$ we perform a quadratic expansion of $1/\lambda_-(E)$ in power of $E+2 \sqrt 2$, yielding $U\simeq c_1 (E+2\sqrt{2})    +c_2 (E+2\sqrt{2})^2$, where 	$c_1=(5+4\sqrt 3-\sqrt{37+16\sqrt 3})/2\simeq 1.942$ and $c_2=\frac{27 (3 - 4 \sqrt 3 + 2 \sqrt {3 (67 - 36 \sqrt 3)})}{\sqrt{134 - 72 \sqrt 3} \left(9 - 2 \sqrt 3 + \sqrt{3 (67 - 36 \sqrt 3)}\right)^2} \simeq 0.365$.
The same result can also be obtained by noting that for small $U$ the dominant contribnution to the  integral in Eq.~(\ref{eq:integ}) corresponds to $\nu=\nu^\prime=-$, leading to a  pole at $E=E_\textrm{ref}$ in the  matrix elements of $R$, while in all the other contributions $E$ can be safely replaced by $E_\textrm{ref}$. 
The binding energy of the pair in the weak coupling regime is then  given by
\begin{equation}\label{eq:FBweakU}
	E_b\simeq -\frac{U}{c_1}+\frac{c_2}{c_1^3}U^2,
\end{equation}
which is shown  in  Fig.~\ref{fig:fig2_EbFBvsU}  by the dot-dashed curve. 
Notice that the linear in $U$ dependence of the binding energy for small $U$, also reported in ~\cite{Torma:PRB2018}, is a direct consequence of the 
localized nature of the single particle states forming the molecule. Indeed, the same behavior was already observed~\cite{Dufour:PRL2012} for two interacting particles in 
the presence of a quasi-periodic lattice, once single-particle localization sets in. 

For strong interactions, we expand $\lambda_{-}(E)$ in power of $1/E$ up to second order. From this we 
	find
\begin{equation}\label{EbstrongUFB}
	E_b\simeq -U-2\sqrt 2 -\frac{6+2\sqrt 5}{U}-\frac{16\sqrt{2/5}+6\sqrt 2}{U^2}.
\end{equation}
The strong coupling expansion (\ref{EbstrongUFB}) is shown in Fig.~\ref{fig:fig2_EbFBvsU}  by the 
dashed curve and agrees well with the exact result for $|U|\gtrsim 5$.  
 Before continuing, it is worth mentioning that the occupation of the two sublattices is asymmetric due to the different connectivity of A and B sites: in the bound state of lowest energy the two constituent particles reside more on the B sublattice, while in the first excited bound state the two particles occupy prevalently the A sites. This point is particularly clear in the strong coupling regime, since the two particles must share the same site to interact.
Expanding the matrix elements in Eq.~(\ref{exact}) to lowest order in $1/E$ yields
\begin{equation}
	R= \begin{pmatrix} 
		1/E+4/E^3 & 4/E^3 \nonumber \\
		4/E^3 & 1/E+8/E^3 \label{eq:mat}
	\end{pmatrix}.
\end{equation}
The eigenvalues of Eq.~(\ref{eq:mat}) are $\lambda_\pm=1/E+ (6\mp 2\sqrt 5)/E^3$ and the associated normalized eigenvectors are, respectively, $\mathbf v_-=(v_1,v_2)$ and $\mathbf v_+=(-v_2,v_1)$,
where $v_1=\sqrt{2/(5+\sqrt 5)}$ and $v_2=\sqrt {(5+\sqrt 5)/10}$.
Hence the probability for the pair to be in the A site is $|v_-^{A}|^2=0.276$ for the ground state 
and $|v_+^{A}|^2=0.724$ for the first excited  bound state. 
This result is consistent with Ref.~\cite{chan2021pairing}, also reporting an asymmetric occupation of the two sublattices for the ground state density profile at finite filling.

\begin{figure}
\includegraphics[width=0.98\columnwidth]{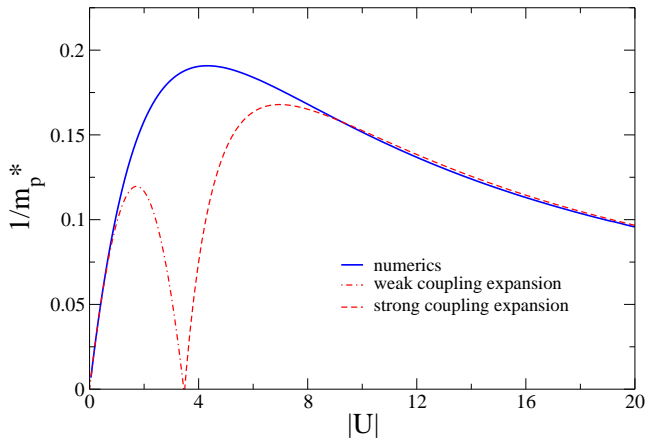}
\caption{(Color online) Inverse effective mass of the deepest bound state at the flat band point as a function of the modulus of the interaction strength. 
The dot-dashed line corresponds to the weak coupling expansion, Eq.~(\ref{eq:m*weakU}), while 
the dashed line represents the strong coupling expansion, Eq.~(\ref{eq:strongmass}).  }
\label{fig:fig2_invmFB}
\end{figure}

%
Let us now discuss the effective mass $m_p^*$ of the pair, which is defined through the relation
$1/m_p^*=E^{\prime \prime} (0)$, where $E(Q)$ is the dispersion relation of the bound state. 
 In Fig.~\ref{fig:fig2_invmFB} we plot the inverse effective mass of the pair  as a function of the 
interaction strength (solid blue line). For $U=-0.1414$ we recover  the numerical result obtained in
Ref.~\cite{Torma:PRB2018}.
In order to derive the  weak coupling expansion for the pair effective mass, we need to calculate the matrix $R$  for a small but finite momentum $Q$. 
To do so, in Eq.~(\ref{eq:integ})  we perform a quadratic expansion in $Q$ for the non singular contributions  and replace therein the energy $E$ by $-2\sqrt 2$.  
The  integration  can then be performed analytically yielding the following approximate expressions for the matrix elements:
\begin{eqnarray}
R_{11}&\simeq &\frac{ (\sqrt 3 -2)(7-\cos Q)+4}{(7-\cos Q) \sqrt 3 (E+2 \sqrt 2)}-\frac{1008+187 Q^2}{1728 \sqrt 6}  \nonumber \\
R_{12}&\simeq  &\frac{2 e^{-i Q/2}\cos (Q/2)}{(7-\cos Q) \sqrt 3 (E+2 \sqrt 2)}+\frac{432-Q(185Q+i 216)}{1728 \sqrt 6} \nonumber\\
R_{22}&\simeq &\frac{4}{(7-\cos Q) \sqrt 3 (E+2 \sqrt 2)} -\frac{5 (144+11 Q^2)}{1728 \sqrt 6}\label{eq:RFBQ}.
\end{eqnarray}
We then substitute the rhs of Eq.~(\ref{eq:RFBQ}) in Eq.~(\ref{det}) and expand  $1/\lambda_{-}=U$ up to  second order in $E+2\sqrt 2$.  We obtain
 $U\simeq f_1(Q) (E+2\sqrt 2)+f_2 (Q) (E+2\sqrt 2)^2 $,
where $f_i(Q)=c_i+d_i Q^2$  are quadratic functions of the momentum. The 
constant $c_i$  are defined as above, while $d_1=\frac{1}{8} \left(\sqrt{\frac{1}{601} \left(1504 \sqrt{3}+3133\right)}-1\right)\simeq 0.261$ and $d_2=\frac{\frac{6}{\sqrt 2} \left(\frac{148042277 -85446942 \sqrt{3}}{\sqrt{67-36 \sqrt{3}}}-5341612 +3094257 \sqrt{3}\right) }{32  \left(36 \sqrt{3}-67\right)^2 \left(9-2 \sqrt{3}+\sqrt{3 \left(67-36 \sqrt{3}\right)}\right)^3}\simeq 0.294$.
Solving for the energy yields $E(Q) \simeq - 2\sqrt 2 +\frac{U}{f_1(Q)}-\frac{f_2(Q)}{f_1(Q)^3}U^2$.  The effective mass of the pair is then given by
\begin{equation}\label{eq:m*weakU}
\frac{1}{m_p^*}\simeq -\frac{2 d_1}{c_1^2}U+\frac{6 c_2 d_1-2 d_2 c_1 }{c_1^4} U^2,
\end{equation}
which is displayed in Fig.~\ref{fig:fig2_invmFB} with the dot-dashed red line.  We see that the inverse effective mass takes its maximum value around 
$U=-4.3$. Interestingly, there is a wide window of $U$ values around this point, where both  the perturbative  
expansion (\ref{eq:m*weakU}) and the strong coupling expansion, that will be derived below 
(see Eq.(\ref{eq:strongmass})), become completely inadequate. In particular,  the pair effective mass 
 is much more sensitive to interband transitions than the binding energy, as one can see by comparing Fig.~\ref{fig:fig2_invmFB} with Fig.~\ref{fig:fig2_EbFBvsU}.

The dependence of $1/m_p^*$  on $|U|$ shown in  Fig.~\ref{fig:fig2_invmFB} is clearly reminescent of the behavior of the superfluid weight $D_s$
investigated in \cite{chan2021pairing}: both quantities scale as $|U|$ for weak interactions and
as $1/|U|$ in the strongly interacting regime. 
For small $|U|$,  the explicit relation between the pair effective mass and the superfluid weight is~\cite{Tovmasyan:PRB2016} 
$D_s=4\pi n(1-n)/m_p^*$, where $n$ is the density (i.e. the number of fermions per lattice site) and  the factor $\pi$ has been added to match the definition of $D_s$ used 
in \cite{chan2021pairing}. In this regime Eq.~(\ref{eq:m*weakU}) yields $1/m_p^*\simeq 0.1385 |U|$, implying 
$D_s\simeq 1.74 |U|n(1-n)$, which is  close to the value $D_s\simeq 1.62 |U|n(1-n)$ found in 
\cite{chan2021pairing} from DMRG data at quarter filling, $n=1/4$.

It is also worth emphasizing that the linear-in-$U$ behavior of the  pair inverse effective mass 
for small $U$   is a generic feature of FB lattices. Indeed, in this regime
 the matrix $R$ in Eq.~(\ref{eq:integ}) takes the approximate form $R\simeq  W/(E-2\varepsilon_\textrm{fb})$, where $\varepsilon_\textrm{fb}$ is the  energy of the flat band and 
 	$W$ is a $Q-$dependent  matrix given by
 \begin{equation}
W=\begin{pmatrix}
	    \sum_q |\alpha_q|^2|\alpha_{Q-q}|^2 &  \sum_q  \alpha_{q}\beta^*_{q} \alpha_{Q-q}  \beta^*_{Q-q} \\ 
 		 \sum_q  \alpha^*_{q}\beta_{q} \alpha^*_{Q-q}  \beta_{Q-q} &  \sum_q |\beta_q|^2|\beta_{Q-q}|^2,
 	\end{pmatrix}  
 \end{equation}
 with $\alpha_k, \beta_k$ being the  amplitude components of the FB states. Therefore the  energy dispersion of the weakly bound state  is $E(Q)=2\varepsilon_\textrm{fb}+U\lambda_W(Q)$, where $\lambda_W(Q)=(W_{11}+W_{22}+\sqrt{(W_{11}-W_{22})^2+4|W_{12}|^2})/2$ corresponds to the largest positive eigenvalue of $W$. In particular the inverse effective mass of the pair is given by $1/m_p^*=U \partial^2 \lambda_W/\partial Q^2$ evaluated at $Q=0$. Notice that this result is completely general, i.e. it does not depend on any assumption of uniform pairing across the two sublattices.

\subsection{Bound states for generic tunneling rates}
\begin{figure}
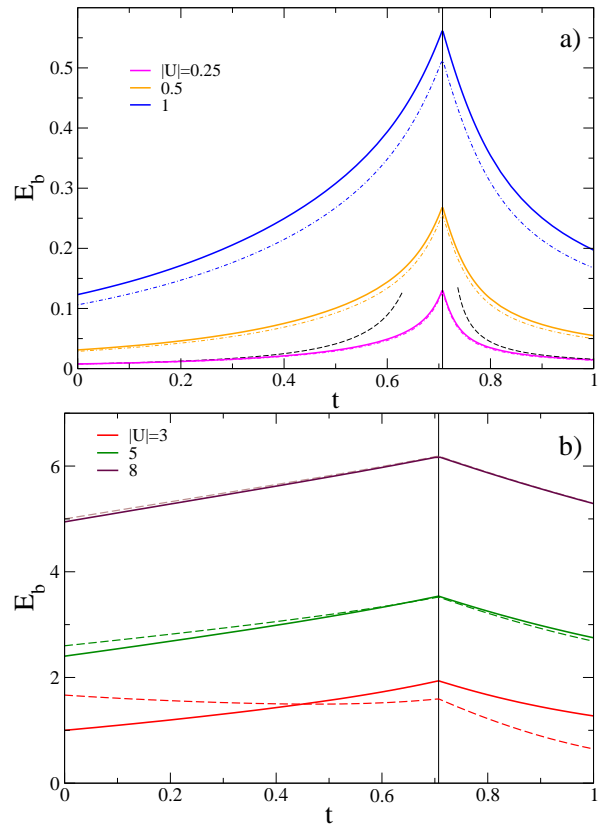

\includegraphics[width=0.9\columnwidth]{figEbvst.eps}
\includegraphics[width=0.9\columnwidth]{figEbstrongU.eps}
\caption{(Color online) a)  Binding energy versus tunneling rate for different values of $U$ in the weakly interacting regime, from 
$U=-0.25$ (bottom) to $U=-1$ (top), shown with solid lines. The dot-dashed curves correspond to the single-band approximation obtained  by  
retaining only the contribution from the lower band in Eq.~(\ref{eq:integ}). The dashed line corresponds to the weak coupling result $E_b=U^2 f$, where $f(t)$ is defined in Eq.~(\ref{eq:f}). Notice that $f$ diverges approaching the flat band point $t=1/\sqrt{2}$  (vertical line). 
b) Same as panel a) but for values of $U$ in the strongly interacting regime, from $U=-3$ (bottom) to $U=-8$ (top),  shown with solid lines. The dashed lines correspond to the strong 
coupling expansion, Eq.~(\ref{eq:apprStrongU2}).
Notice that the binding energy takes its maximum value at the flat band point for any value of $U$. 
}
\label{fig:fig1_Ebvst}
\end{figure}

We investigate here the properties of the lowest energy bound state in the absence of the flat band, i.e. for an arbitrary $t\neq 1/\sqrt 2$. From Eq.~(\ref{disp}) we find that 
the reference energy is given by
\begin{equation}
E_\textrm{ref}=
\begin{cases}
 2 (t - \sqrt{4 + t^2})& \text{if $t<\frac{1}{\sqrt{2}}$} \label{eq:Eref}\\
 -4t &  \text{if $t>\frac{1}{\sqrt{2}}$}.
 \end{cases}
\end{equation}
This quantity exhibits a maximum with discontinuous derivative at the FB point, due to the crossing between the two defining functions in Eq.(\ref{eq:Eref}).
In Fig.~\ref{fig:fig1_Ebvst} we plot the binding energy as a function of the tunneling rate $t$, for different values of $U$ (solid lines). The two panels (a) and (b) correspond to the weak and the strong coupling  regimes, respectively.
We see that $E_b$ takes its maximum value in correspondence of the FB point (solid vertical line), for all values of the interaction
strength.
The origins of the peak in the weak and in the strongly interacting regimes are different. In the first case, it directly follows from the fact that at the FB point $E_b$ 
scales linearly in $U$, while  for any other values of $t$ the binding energy  grows only quadratically in $U$.  
 To see this, we  perform a quadratic expansion 
around the bottom of the  lower band, $\epsilon_{-}(q)\simeq \epsilon_{-}(q_\textrm{B})+ \epsilon_{-}^{\prime \prime}(q_\textrm{B})(q-q_\textrm{B})^2/2$. 
Next, we approximate the numerator in the rhs of  Eq.~(\ref{eq:integ}) by a constant, $M_{--}(q,-q)  \simeq M_{--}(q_\textrm{B},-q_\textrm{B})$ 
and integrate over momentum. From Eq.~(\ref{det}) we obtain
\begin{equation}\label{eq:apprWeakU}
\frac{1}{U}\simeq -\frac{(|a_{q_\textrm{B} -}|^4+|b_{q_\textrm{B} -}|^4)}{ 2\sqrt{ \epsilon_{-}^{\prime \prime}(q_\textrm{B})}\sqrt E_b},
\end{equation}
 showing that for small $U$  the binding energy grows as $U^2$. This behavior is well known 
  from the linear chain limit $t=0$, where $E_b=\sqrt{U^2+16}-4$.
 Eq.~(\ref{eq:apprWeakU})  breaks down for $t=1/\sqrt 2$, 
since the single-particle  inverse effective mass  vanishes at the FB point.
 We therefore write $E_b\simeq U^2 f(t)$, where 
\begin{equation}\label{eq:f}
 f(t)=\begin{cases}
 -\frac{(2+t^2)^2 \sqrt{4+t^2} (2+t^2-t \sqrt{4+t^2})^2  }{(4+t^2-t \sqrt{4+t^2})^4  (-1+t(-t+\sqrt{4+t^2}))} & \text{if $t<\frac{1}{\sqrt{2}}$} \\
\frac{ t}{8 t^2-4} &  \text{if $t>\frac{1}{\sqrt{2}}$}.
 \end{cases}
\end{equation}
 is a function of the tunneling rate, which is obtained by substituting 
in Eq.~(\ref{eq:apprWeakU}) the explicit expressions for the effective mass and the  amplitudes 
of the Bloch states, given in Eq.(\ref{eq:m*}) and  Eq.~(\ref{fun}), respectively.  
In Fig.~\ref{fig:fig1_Ebvst} (a) we display  the result based on Eq.~(\ref{eq:f}) for the weakest interaction considered, $U=-0.25$ (dashed line).  We see that there is a wide region of $t$ values  around the FB point, where the  weak coupling 	approximation (\ref{eq:apprWeakU}) deviates significantly from the numerical result.
We emphasize that Eq.~(\ref{eq:apprWeakU})  relies on the assumption that $E_b\ll w$, where $w=|\sqrt{4+t^2}-3t |$ is the width of the lowest energy band. This condition is necessarily violated near the FB point, where the bandwidth  vanishes. 

\begin{figure}
	\includegraphics[width=0.98\columnwidth]{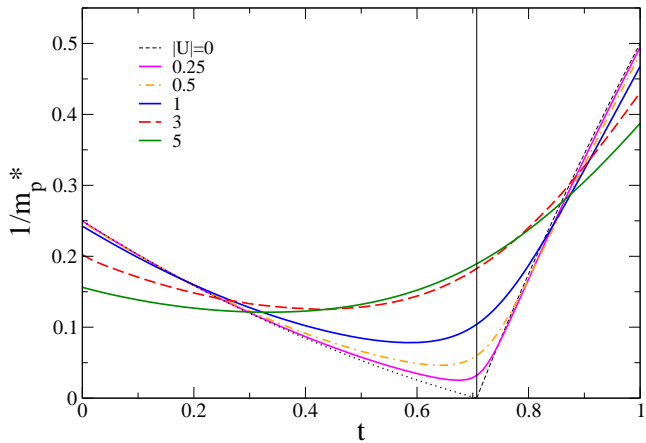}
	\caption{(Color online) Inverse effective mass of the deepest bound state as a function of the tunneling rate $t$ for different values of the interaction strength $U=-0.25$ (magenta solid line), $U=-0.5$ (orange dot-dashed line), $U=1$ (blue solid line), $U=-3$ (red long dashed line) and $U=-5$ (green solid line). The dotted line corresponds to the limit of vanishing attractive interactions, $m_p^*=2m^*$. The vertical line represents the flat band point $t=1/\sqrt{2}$.}
	\label{fig:fig2_invmvst}
\end{figure}

The dot-dashed curve  in   Fig.~\ref{fig:fig1_Ebvst} (a) refers to the single-band approximation  for the binding energy, obtained by neglecting completely the  upper band  in Eq.~(\ref{eq:integ}), thus retaining only the contribution  corresponding to $\nu=\nu^\prime=-$. For  weak interactions the approximation is  
	accurate  for any value of $t$, in stark contrast with the weak coupling expansion (\ref{eq:apprWeakU}), pointing out that all momenta inside the Brillouin zone must be taken into account when approaching  the FB point. As $|U|$ increases, interband transitions become important and the 
single-band approximation deviates more and more from the exact numerics.

In the presence of a very strong attraction, the two particles sit at the same lattice site and form a tightly bound state. Since $E\sim U$ is large and negative, 
the binding energy  reduces to $E_b \simeq  -U+E_\textrm{ref}$. Thus, in this regime  the binding energy 
simply mirrors the reference energy, showing a singular peak for  $t=1/\sqrt 2$, as displayed in Fig.~\ref{fig:fig1_Ebvst} (b).
In order to include higher order corrections to the binding energy,
we use the formula $\frac{1}{(E-x)}=\frac{1}{E} \sum_{n=0}^{+\infty} (\frac{x}{E})^n $ in the rhs of 
Eq.~(\ref{eq:integ}), with $x=\varepsilon_\nu +\varepsilon_{\nu^\prime}$, and cut the series after the $n=3$ term. The integration over momentum can then be done analytically and from Eq.~(\ref{det}) we obtain 
\begin{eqnarray}
\frac{1}{U}&\simeq &\frac{E^3+12 t +4 E (1+t^2)}{E^4}\nonumber\\
&&-\frac{4 \sqrt{13t^2+E(6t+4t^3)+E^2(1+t^4)}}{E^4}.\label{eq:apprStrongU} 
\end{eqnarray}
Expanding the rhs of Eq.~(\ref{eq:apprStrongU})  in power of $E$, up to second order, gives $U\simeq E+a_1/E+a_2/E^2$, where 
$a_1=-4(1+t^2+\sqrt{1+t^4})$ and $a_2=-12t-(12t+8t^3)/\sqrt{1+t^4}$.
From this we obtain 
\begin{equation}\label{eq:apprStrongU2}
E_b \simeq  -U+E_\textrm{ref}-\frac{a_1}{U}-\frac{a_2}{U^2},
\end{equation} 
which reduces to Eq.~(\ref{EbstrongUFB}) for $t=1/\sqrt2$.
The strong coupling prediction   (\ref{eq:apprStrongU2}) is displayed in Fig.~\ref{fig:fig1_Ebvst} (b) with dashed lines. We see that the approximation works better and better as $|U|$ increases.  

Let us now discuss the behavior of the two-body inverse effective mass.
In Fig.~\ref{fig:fig2_invmvst} we display our numerical results as a function of the tunneling rate $t$ and for different values of the interaction strength. The dotted line corresponds to the noninteracting limit, where the bound state breaks down and the pair effective mass reduces to
twice the single-particle mass, $m_p^*=2m^*$.
We see   that, far from the FB point, weak interactions tend to slightly increase the effective mass of the pair.
In contrast, close to it, the  inverse mass is strongly enhanced by interactions, an effect which persists until $U\approx -5$. 
We also note that the minimum in the inverse effective mass shifts towards smaller values of $t$ as $|U|$ increases.

The interaction  correction to $1/m_p^*$  in the weak coupling regime can be obtained by generalizing Eq.~(\ref{eq:apprWeakU}) to a finite 
momentum of the pair. In particular, for $Q>0$ the dominant contribution to the integral in  Eq.~(\ref{eq:integ}) comes from the region centered around 
$q=q_B+Q/2$, with $q_B$ defined as above. By replacing $q_\textrm{B}$ with $q_\textrm{B}+Q/2$ in Eq.~(\ref{eq:apprWeakU}) and taking the square of both sides of
the equation, we obtain the  dispersion relation of the bound state
\begin{equation}\label{auxM}
E(Q)= 2\epsilon_{-}(q_B+Q/2)-\frac{(|a_{q_\textrm{B}+Q/2 -}|^4+|b_{q_\textrm{B}+Q/2 -}|^4)^2}{4 \epsilon_{-}^{\prime \prime}(q_\textrm{B}+Q/2)}U^2.
\end{equation}
From Eq.~(\ref{auxM}) we then find that the effective mass of the pair in the weak coupling regime reduces to
\begin{equation} \label{eq:m*weakGent}
\frac{1}{m_p^*}\simeq \frac{1}{2m^*}+h U^2,
\end{equation}
where  $h$ is a function of the tunneling rate, whose explicit form is obtained by 
making use of Eqs~(\ref{disp}) and (\ref{fun}) in Eq.~(\ref{auxM}). We find $h(t)=\frac{4 - t (8 u + 
	t (60 + t (40 u + 
	t (86 + t (25 u + t (36 + 5 t (t + u)))))))}{16 u^5  (2 t^2-1)}$
for $t<1/\sqrt 2$, with $u=\sqrt{4+t^2}$, while $ h(t)=(1+t^2)/(16t(1-2t^2))$ for  $t>1/\sqrt 2$.
Notice that $h$ diverges for $t=1/\sqrt 2$ , since in  Eq.~(\ref{auxM})  $\epsilon_{-}^{\prime \prime}(q)=0$ in the entire Brillouin zone. The divergence signals that 
Eq.~(\ref{eq:m*weakGent}) does not hold at the FB point, in agreement with Eq.~(\ref{eq:m*weakU}).

The strong coupling expansion for the pair effective mass can be obtained by following the same procedure used to 
derive Eq.~(\ref{eq:apprStrongU}), but this time we retain the full $Q$ dependence of the matrix elements in the rhs of Eq.~(\ref{eq:integ}). The integration 
can still be performed analytically and from Eq.~(\ref{det}) we obtain
\begin{widetext}
\begin{equation}\label{eq:apprStrongUQ}
\frac{1}{U}\simeq \frac{1}{E^4}  (E(4+E^2)+6t(1+\cos Q)+ 2Et^2(1+\cos Q)-\sqrt{8 (E+3t)^2(1+\cos Q) +4 t^2(1+E t+(3+Et)\cos Q)^2}),
\end{equation}
\end{widetext}
which provides an implicit equation for the dispersion relation of the bound state. To make it explicit, 
we expand the rhs of 
 Eq.~(\ref{eq:apprStrongUQ}) in power of $1/E$, retaining  up to second order terms, and solve for the energy. 
This yields
\begin{eqnarray}
\frac{1}{m_p^*} &\simeq &-\frac{1+2 t^2(t^2+\sqrt{1+t^4} )}{U\sqrt{1+t^4}} \nonumber \\
&-&\frac{t(3+8t^2+6t^6+6\sqrt{1+t^4}+6t^4 \sqrt{1+t^4})}{U^2 (1+t^4)^{3/2}}, \label{eq:strongmass}
\end{eqnarray}
holding for any value of $t$, included the FB point (see the dashed curve  in Fig.~\ref{fig:fig2_invmFB} ). Notice that  the $1/U^2$ correction in Eq.~(\ref{eq:strongmass}) accounts for the non-monotonic behavior of the inverse effective mass displayed in Fig.~\ref{fig:fig2_invmvst}, including 
the shift of the minimum towards smaller values of $t$ as interaction effects become stronger. 

\section{BCS-BEC crossover}
\label{sec:manybody}
In this section we use the DMRG method to  investigate the ground state properties of a spin-1/2 Fermi gas on the sawtooth lattice undergoing the BCS-BEC crossover. 
The underlying Fermi-Hubbard Hamiltonian is given by
\begin{eqnarray}\nonumber
	H&=&\sum_{i\alpha} [t c_{i\alpha}^{B\dagger} c_{i+1\alpha}^B+    t^\prime c_{i\alpha}^{A\dagger} (c_{i\alpha}^B+c_{i+1\alpha}^B) + \textrm{h.c.}]\\
	&+&U \sum_{i} (n_{i\uparrow}^A n_{i\downarrow}^A+n_{i\uparrow}^B n_{i\downarrow}^B),
	\label{eq:Hbis}
\end{eqnarray}
where  $c_{i\alpha}^{\sigma \dagger} (c_{i\alpha}^\sigma)$ is the local
creation (annihilation) operator for fermions with spin component ${\alpha=\uparrow,\downarrow}$ in the sublattice $\sigma=A, B$, and  $n_{i\alpha}^\sigma=c_{i\alpha}^{\sigma\dagger} c_{i\alpha}^\sigma$ are the corresponding density operators. 
We recall  that $t^\prime=1$ in our energy units. 
We define the density  of the two spin components with respect to the total number of lattice sites, $n_\alpha=N_\alpha/(2L)$, where $N_\alpha$ is the number of fermions with spin $\alpha$. 
In this work we  restrict our attention to fully paired systems, corresponding to equal densities of the two spin components, and assume that only the flat band is occupied in the absence of interactions, that is
$n_\uparrow=n_\downarrow<1/2$.

Two important observables characterizing the BCS-BEC crossover in Fermi gases are  the pairing gap $\Delta_\textrm{pg}$ and the chemical potential $\mu$. The first, also known as the spin gap,
corresponds to the energy needed to break a pair in the many-body system by reversing one spin, while the second corresponds to half  the energy change upon adding a pair (one fermion with spin up and one fermion with spin down) to the system.
Let $\epsilon(n,s)$ be the ground state energy per lattice site,
expressed in terms of the total fermion density $n=n_\uparrow+n_\downarrow$, and the spin density $s=n_\uparrow- n_\downarrow$.
The pairing gap and the chemical potential are given by  
\begin{equation}\label{DeltaANDmu}
\mu=\left(\frac{\partial \epsilon}{\partial n}\right)_{s=0},\;\;  \Delta_\textrm{pg}=2 \left(\frac{\partial \epsilon}{\partial s}\right)_{s=0}.
\end{equation}
We compute the ground state energy $E(N_\uparrow, N_\downarrow)$ of the system as a function of the spin populations $N_\alpha$ for a large enough system size $L$. 
We consider systems sizes up to $L=60$, corresponding to $120$ sites, with open boundary conditions.  
 
 We evaluate the chemical potential by approximating the derivative in
 Eq.(\ref{DeltaANDmu}) by a finite difference, $\mu \simeq (E(N_\uparrow +1,N_\downarrow +1)-E(N_\uparrow,N_\downarrow))/2$.  For the pairing gap, 
we  use    $-\Delta_\textrm{pg} \simeq  E(N_\uparrow +1,N_\downarrow +1)-2 E(N_\uparrow +1,N_\downarrow) +E(N_\uparrow,N_\downarrow)$.
In the thermodynamic limit, this  formula is equivalent to the finite difference  $\Delta_\textrm{pg} \simeq  E(N_\uparrow +1,N_\downarrow -1)-E(N_\uparrow,N_\downarrow)$, but is 
less sensitive to finite-size effects.
For vanishing densities, both the pairing gap and the chemical potential possess a well defined limit, which is consistent with the solution of  the two-body problem.
The pairing gap reduces to the binding energy, since for $N_\uparrow=N_\downarrow=0$ we find from Eq.~(\ref{DeltaANDmu}) that
$\Delta_\textrm{pg} =E(2,0)-E (1,1)=E_\textrm b$. Here we use the fact that $E(2,0) =2E(1,0)=E_\textrm {ref}$ and $E(1,1)=E(Q=0)$, where $E(Q)$ is the energy dispersion of the  two-body bound state calculated in Sec.\ref{sec:boundstates}.
 From Eq.~(\ref{DeltaANDmu}) we instead find that $\mu=E(1,1)/2$, since  $E(0,0)=0 $, implying that the chemical potential reduces to
 \begin{equation}\label{muSC}
 \mu=\frac{-E_\textrm b +E_\textrm{ref}}{2}.
 \end{equation} 
 A peculiar feature of 1D fermionic systems is that interaction effects become stronger as the density $n$ decreases. As a consequence, the binding energy provides an upper bound 
 for the pairing gap. In contrast, the two-body prediction (\ref{muSC})  is a lower bound for the chemical potential, because the inverse compressibility $\partial \mu/\partial n$ must be positive or null to ensure the energetic stability of the gas.
 
Before presenting our results, we  emphasize that the pairing gap discussed here is different from the pairing order parameters $\Delta^{\sigma}$ investigated in Ref.~\cite{chan2021pairing}. The latter are defined in terms of the diagonal part of the sublattice-resolved pair-pair correlation function through the relation $|\Delta^\sigma|^2/U^2 = \sum_i (\langle c_{i\uparrow}^{\sigma\dagger} c_{i\downarrow}^{\sigma\dagger}  c_{i\downarrow}^\sigma c_{i\uparrow}^ \sigma \rangle -\langle n_{i\uparrow}^\sigma \rangle \langle n_{i\downarrow}^\sigma \rangle)/L$. These quantities clearly depend on the many-body wave-function and  can therefore take different values on the two sublattices, $\Delta^{A}\neq \Delta^B$, due to the different connectivity of A and B sites.
In contrast, the pairing gap is obtained solely from the ground state energy of the 
system through Eq.~(\ref{muSC})  and therefore cannot depend on the sublattice index.
 It is also worth adding that in Ref.~\cite{chan2021pairing} the pairing parameters $\Delta^{\sigma}$  are shown to be increasing functions of  the density, while the pairing gap discussed here exhibits the opposite behavior (see Fig.~\ref{fig:gapvsdensity} (b) below). Notice that the same difference in  the density dependence of the two observables is also present in the linear chain limit $t=0$, see for instance~\cite{Marsiglio:PRB1997}.

\subsection{Results at flat band point}
\begin{figure}
\includegraphics[width=0.98\columnwidth]{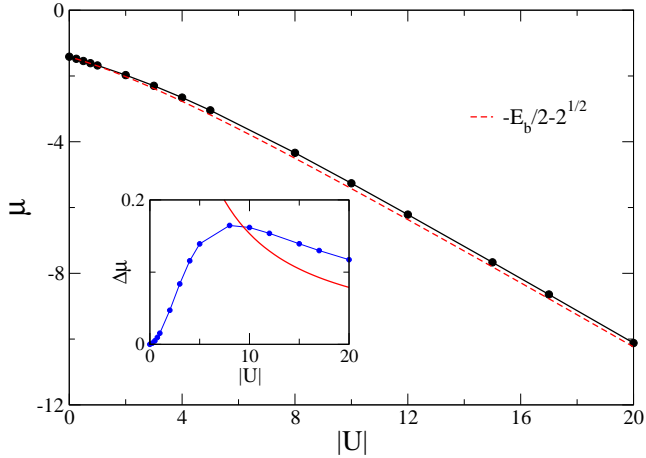}
\caption{(Color online) Chemical potential  at the flat band point  as a function of the modulus of the interaction strength  for filling $n_\uparrow=n_\downarrow=1/3$ (black circles). 
The dashed line represents the zero density limit of the chemical potential, $\mu(n=0)=-E_b/2+E_\textrm{ref}/2$, see Eq.(\ref{muSC}). 
The data symbols in the inset correspond to the difference $\Delta \mu=\mu-\mu(n=0)$,  plotted as a function of $|U|$. The red solid curve shows the corresponding result for $t=0$, obtained from 
the Bethe ansatz integral equations, where $\Delta \mu=1$ in the noninteracting limit. 
The solid connecting lines are a guide to the eye.}
\label{fig:muvsU}
\end{figure}
In   Fig.~\ref{fig:muvsU}  we plot the chemical potential versus $|U|$ at the FB point, together with the zero density prediction (\ref{muSC}). 
For weak enough interactions, finite density corrections are small, due to the infinite compressibility associated to the flat band. 
 From Eqs~(\ref{eq:FBweakU}) and (\ref{muSC})  we obtain, 
to first order in $U$, $\mu\simeq -\sqrt 2 +U/(2c_1)$, which is fully consistent with our numerics. This result differs from the BCS mean field estimate given in~\cite{chan2021pairing},
where the  linear in $U$ correction to the chemical potential was found to  explicitely depend  on the density.

\begin{figure}
	\includegraphics[width=0.98\columnwidth]{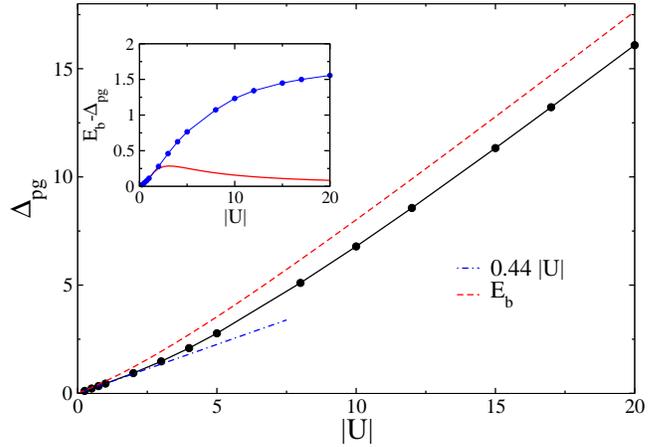}
	\caption{(Color online) Pairing  gap  at the flat band point as  a function of the modulus of the interaction strength, for filling $n_\uparrow=n_\downarrow=1/3$ 
		(black circles). The dashed line represents the zero density 
		limit of the pairing gap, $\Delta_\textrm{pg}(n=0)=E_b$. The data symbols in the inset correspond to the difference $E_b-\Delta_\textrm{pg}$, plotted as a function of $|U|$. The red solid line is the corresponding prediction for $t=0$, obtained from the Bethe ansatz integral equations, showing that the difference instead vanishes as $1/|U|$ for strong interactions. 
        The connecting solid lines are a guide to the eye.}
\label{fig:gapvsU}
\end{figure}

In the inset of Fig.~\ref{fig:muvsU}  we plot the difference between the two curves in the main panel, corresponding to $\Delta \mu =\mu-\mu(n=0)$.
 This quantity shows a non-monotonic behavior as a function of $|U|$: $\Delta \mu=0$ in the absence of interactions, then it increases with $|U|$, reaching  a maximum around  $|U|\approx 8$, and finally decreases as $\Delta \mu \sim 1/|U|$ in the strong coupling regime.   Notice that $\Delta \mu \geq 0$, because the inverse compressibility $(\partial \mu/\partial n)_{s=0}$  must be positive or zero to ensure the mechanical stability of the system.
For large $|U|$  bound states behave as point-like hard-core bosons, hopping between neighboring sites of the sawtooth lattice  and experiencing repulsive nearest-neighbor interactions as well as a uniform potential of different strength in the two sublattices. Since all these processes are characterized by the same energy
scale $1/U$, as demonstrated  in Ref.\cite{Micnas:RMP1990},  the leading density correction to Eq.~(\ref{muSC}) must be of the same order.
For comparison, in the inset of Fig.~\ref{fig:muvsU}   we also show  the corresponding result for the integrable point
	$t=0$ (red solid curve). This is obtained by solving numerically the Bethe ansatz integral equations, as  done in Ref~\cite{Heidrich:PRA20210}. 
Differently from the FB case,  at the integrable point $\Delta \mu$ is a monotonic decreasing  function of $|U|$. In particular   
$\Delta \mu=1$ for $U=0$ (because $n=2/3$), while in the  strong coupling regime we find $\Delta \mu \sim1/|U|$. 

In Fig.~\ref{fig:gapvsU} we show the pairing gap as a function of the modulus of the interaction strength (black circles), 
together with the two-body binding energy (red dashed line).
 For weak interactions the numerical data are well fitted by $\Delta_\textrm{pg}=0.44 |U|$, shown by the blue dot-dashed line. For strong interactions, the difference between the binding energy and the pairing gap saturates to a constant value, as shown in the inset of Fig.~\ref{fig:gapvsU}. For comparison, in the inset we also display the corresponding prediction for $t=0$ (red solid line), showing that the difference is instead non-monotonic and decreases as $1/|U|$  in the strong coupling regime. 
 
\begin{figure}
\includegraphics[width=0.98\columnwidth]{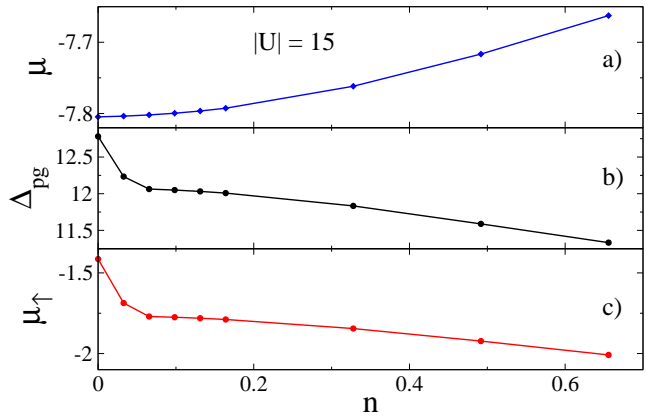}
\caption{ Chemical potential (a), pairing gap (b) and excess energy (c)  at the flat band point
as a function of the  total fermion density $n$, for $U=-15$. 
The solid connecting lines are a guide to the eye.}
\label{fig:gapvsdensity}
\end{figure}

The chemical potential and the pairing gap at the FB point show very different behaviors as a function of the 
density, as outlined in Fig.~\ref{fig:gapvsdensity} (panels a and b) for $U=-15$. While the chemical potential is nearly constant for small $n$,  the
pairing gap decreases very rapidly at low densities $(n \lesssim 0.08)$, suggesting a singular (i.e. non-analytical) behavior for $n=0$. Moreover finite-density effects  are typically one order of magnitude larger  for the pairing gap than for the chemical potential.  
The above results strongly contrast with  the known behavior  at the  integrable point $t=0$, where~\cite{Woynarovich:1991} 
\begin{equation}\label{eq:exact}
	\mu= -\frac{\sqrt{U^2+16}}{2}+\frac{\pi^2 n^2}{4\sqrt{U^2+16}},\;\;  \Delta_\textrm{pg}=E_b-\frac{\pi^2 n^2}{2\sqrt{U^2+16}},
\end{equation}
which is valid for $n\ll 1$ and $|U|\gg 1$, with $E_b=\sqrt{U^2+16}-4$. From Eq.~(\ref{eq:exact}) we 
infer  that density corrections are of the same order $(1/|U|)$ for both quantities and no singular behavior 
occurs at zero density.

 To better understand the origin of the strong finite-density effects on the pairing gap, we study the \emph{excess energy} $\mu_\uparrow$, 
 corresponding to  the change
 in the ground state energy of the system upon adding an extra spin up fermion, 
  $\mu_\uparrow=E(N_\uparrow +1,N_\downarrow) -E(N_\uparrow,N_\downarrow)$. 
  From Eq.(\ref{DeltaANDmu}) we find that this quantity  is  related to the previous observables by the general equation
\begin{equation}\label{eq:mup}
\mu_\uparrow=\mu+ \Delta_\textrm{pg}/2,
\end{equation} 
 holding for any tunneling rate $t$ and interaction strength $U$. 
From  Eq.~(\ref{eq:mup})  we then find
 \begin{equation}\label{asiU}
 	\Delta_\textrm{pg}= E_b-E_\textrm{ref}+2\mu_\uparrow-2 \Delta \mu,
 \end{equation}
implying that the density dependence of the paring gap comes not only from the equation of state, as in 
Eq.~(\ref{eq:exact}), but also from the excess energy. This point is particularly clear in Fig.~\ref{fig:gapvsdensity} (c), where $\mu_\uparrow$ is plotted as a function of the density, showing that the
excess energy is responsible for the anomalous behavior of the pairing gap at low density.

\begin{figure}
	\includegraphics[width=0.98\columnwidth]{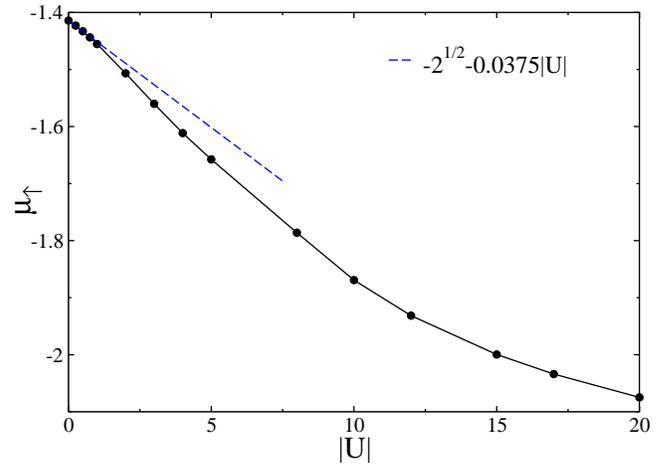}
	\caption{(Color online) Excess energy  $\mu_\uparrow$ (see text for definition) at the flat-band point 
		as a function of $|U|$  for filling
		$n_\uparrow=n_\downarrow=1/3$ (black circles). The blue 
		dashed line represents the asymptotic behavior for weak interactions. 
		The connecting line is a guide to the eye.}
	\label{fig:mupvsU}
\end{figure}

In Fig.~\ref{fig:mupvsU} we plot the excess energy   as a function of the interaction strength. We see that $\mu_\uparrow$ is a decreasing function of $|U|$.
For a noninteracting gas $\mu_\uparrow=\mu=\epsilon_F=-\sqrt 2$, since for $n<1$ the upper dispersive band is empty. To first order in $U$ we find 
 $\mu_\uparrow\simeq -\sqrt 2 + U F(n)$, where $F(n)$ is a  function of the density, satisfying 
 $F(n=0)=0$. This behavior is shown in Fig.~\ref{fig:mupvsU} by the blue dashed line for $n=2/3$. 
 From Eq.~(\ref{asiU}) we then find, to the same order, that $ \Delta_\textrm{pg}\simeq U(2F(n)-1/c_1)$ since $\Delta \mu\simeq 0$. 
For large $|U|$ the excess energy does not scale linearly with $U$, as the chemical potential does, because adding an extra fermion 
to a fully paired system does not change the number of pairs.   Instead  it saturates to a 
density-dependent value, which sits well below the energy of the flat band for the chosen density, $\mu_\uparrow < E_\textrm{ref}/2$.
From Eq.~(\ref{asiU}) this implies that the pairing gap is strongly reduced  by the finite density
as shown in the inset of Fig.~\ref{fig:gapvsU}, while corrections from the equation of state are subleading, since $\Delta \mu \sim 1/|U|$.

It is interesting to note that  for strong interactions Eq.~(\ref{asiU}) reduces to $ \Delta_\textrm{pg}\simeq -E(1,1)+2\mu_\uparrow$, showing that the pairing gap yields the ground state energy of the pair in vacuum, but measured with respect to the \emph{many-body} reference energy  $2 \mu_\uparrow$, instead of the  reference energy $E_\textrm{ref}$.  
In particular, the condition $\mu_\uparrow < E_\text{ref}/2$ indicates that  the excess fermion and tightly bound pairs tend to attract each other, possibly leading to the formation of three-body bound states, as discussed in Sec.~\ref{sec:3body} below. We stress that this effective attraction is instead absent at the integrable point  $t=0$, since  for large $|U|$ Eq.~(\ref{eq:exact}) and Eq.(\ref{eq:mup}) yield  $\mu_\uparrow=-2=E_\text{ref}/2$.

\subsection{Results for generic tunneling rate}

\begin{figure}
\includegraphics[clip,width=0.98\columnwidth]{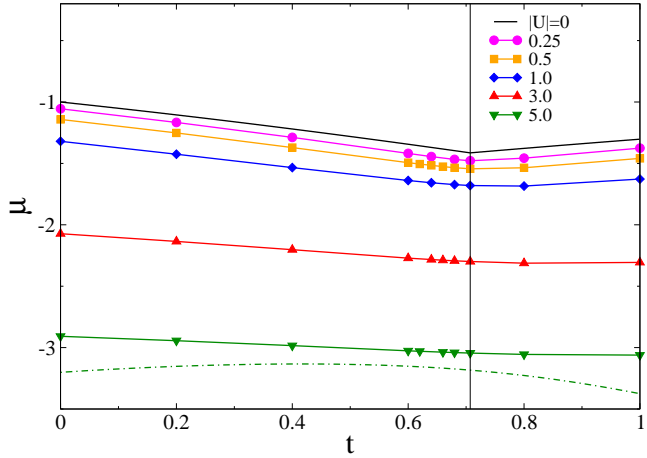}
\caption{(Color online)  Chemical potential $\mu$ as a function of the tunneling rate $t$ plotted for  different values of $U=0, -0.25, -0.5, -1,-3, -5$ with $n_\uparrow=n_\downarrow=1/3$. 
The two-body limit $\mu=(-E_\textrm b +E_\textrm{ref})/2$ for $U=-5$ is shown with the dot-dashed line. The vertical solid line indicates the flat-band point  $t=1/\sqrt{2}$. The connecting lines are a guide to the eye.}
\label{fig:muvst}
\end{figure}

In Fig.~\ref{fig:muvst}  we plot the chemical potential as a function of the tunneling rate, for a
fixed total density $n=2/3$; the different curves correspond to 
different values of $|U|$. In the noninteracting limit (black solid line), the chemical potential coincides with the Fermi energy $\varepsilon_F$ of the system, implying that
 $\mu=\varepsilon_-(q_\textrm{B}-\pi n/2)$. The curve exhibits a minimum at the FB point, due to the moderately large value of the density. 
In the presence of interactions, however, this minimum progressively disappears and the chemical potential  flattens out because $\mu \approx U/2$ for large $|U|$.  
In  Fig.~\ref{fig:muvst}  we also display the zero density limit (\ref{muSC})  of the chemical potential for $U=-5$. We see that the system is more compressible at $t\approx 0.6$.

 \begin{figure}
\includegraphics[clip,width=0.98\columnwidth]{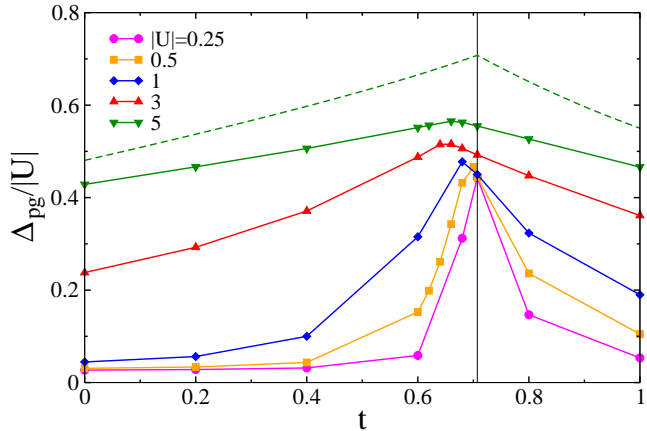}
\caption{(Color online)  
Ratio between the pairing gap and $|U|$, plotted  as a function of the tunneling rate $t$  for different values of $U=-0.25, -0.5, -1,-3, -5$ with  $n_\uparrow=n_\downarrow=1/3$.  
The two-body  binding energy $E_b$ for 
$U=-5$ is shown for reference (dashed line).
The vertical solid line indicates the flat-band point  $t=1/\sqrt{2}$. The connecting lines are a guide to the eye.
}
\label{fig:gapvst}
\end{figure}

In Fig.~\ref{fig:gapvst} we plot the ratio $\Delta_\textrm{pg}/|U|$ between the spin-gap and the modulus of the interaction strength
as a function of the tunneling rate $t$ for increasing values of $|U|$.
The obtained results are clearly similar to their two-body counterpart, presented in Fig.~\ref{fig:fig1_Ebvst}, 
showing a drastic enhancement of pairing in the vicinity of the FB point for weak to moderate interactions. 
There are other interesting and unexpected effects brought about by the finite  density.
 First,  the position of the maximum of the pairing gap drifts to smaller values of $t$
as $|U|$ increases, while the two-body binding energy  always remains peaked at $t=1/\sqrt 2$ (see Fig.~\ref{fig:fig1_Ebvst}). 
Second, for strong interactions density corrections are more prominent near the FB point, as can be seen in  Fig.~\ref{fig:gapvst}, where we contrast the pairing gap with the binding energy (dashed line) for $U=-5$.
By comparing Fig.~\ref{fig:muvst} with Fig.~\ref{fig:gapvst}, we also see that 
 density corrections for the chemical potential can be significantly smaller than for  the pairing gap, except in a neighborood of the integrable point $t=0$, where Eq.~(\ref{eq:exact}) applies.
 
\begin{figure}
\includegraphics[width=0.98\columnwidth]{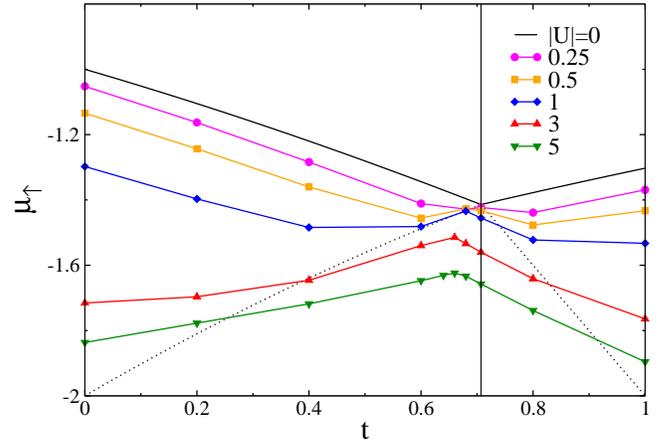}
\caption{(Color online)  
Excess energy versus tunneling rate   for different values of $U=0, -0.25, -0.5, -1,-3, -5$ with  $n_\uparrow=n_\downarrow=1/3$. The dotted line represents the energy bottom $E_\text{ref}/2$ of the lower band.
The vertical solid line indicates the flat-band point  $t=1/\sqrt{2}$. The connecting lines are a guide to the eye.}
  	\label{fig:mupvst}
\end{figure}

In  Fig.~\ref{fig:mupvst} we plot the corresponding results for the excess energy  as a function of the tunneling rate.  
 Far from the FB point, weak interactions cause a fast decrease of $\mu_\uparrow$ with respect to the Fermi energy, while  
 near the FB point the decrease is rather modest.  As a consequence, a local maximum appears, 
 drifting towards smaller values of $t$ as $|U|$ increases and turning into a global maximum at $t\approx 0.66$.   
 Since for finite interactions the chemical potential in Fig.~\ref{fig:muvst} depends smoothly on the tunneling rate,
 we find from  Eq.~(\ref{eq:mup}) that the drift of the peak in the pairing gap simply reflects the behavior of the excess energy. 
As $|U|$ increases, we see from Fig.~\ref{fig:mupvst}  that there is a growing window of  $t$ values around the FB point, in which  the condition $\mu_\uparrow < E_\text{ref}/2$ is satisfied. In this region, the pairing gap is density-depleted for arbitrary large $|U|$, as shown in Fig.~\ref{fig:gapvst}.  The
 anomalous attraction between Cooper pairs and extra fermions is therefore  not specific to the FB point, but appears as a general feature of  mutiband lattices as opposed to linear chains, provided $|U|$ is large enough.

\section{Three-body bound states}
\label{sec:3body}

\begin{figure}
	\includegraphics[width=0.98\columnwidth,clip]{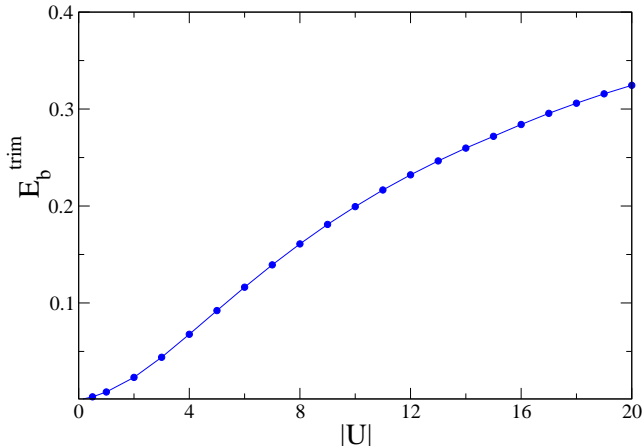}
	\caption{(Color online)  
		Binding energy of a three-body bound state (trimer) at the flat band point $t=1/\sqrt{2}$ plotted as a function of the modulus of the  interaction strength. The connecting line is a guide to the eye. The binding energy saturates to a constant $C\simeq 0.531$ for $|U|=\infty$.}
	\label{fig:trimers}
\end{figure}

In this section we  consider two spin up fermions and one spin-down fermion, obeying the Hamiltonian 
(\ref{eq:Hbis}), and show that  the fermion-pair anomalous attraction can induce  the formation of a three-body bound state  in vacuum (see ~\cite{Mattis:RMP1986,Burovski:PRL2009,Orso:PRL2010,Valiente:PRA20103b,Dalmonte:PRL2011} for earlier studies of trimers in 1D fermionic or bosonic lattice models).
The binding energy $E_b^\textrm{trim}$ of the trimer is  defined as
\begin{equation}\label{def:Ebtrim}
	E_b^\textrm{trim}=-E(2,1)+E(1,1)+E(1,0),
\end{equation}
under the assumption  that  the length $L$ of the chain is  infinite. 
We calculate the ground state energy $E(2,1)$ of the system  numerically, based on the DMRG method,
and extract $E_b^\textrm{trim}$ from Eq.~(\ref{def:Ebtrim}). 
The obtained results for $t=1/\sqrt 2$ are displayed in Fig.~\ref{fig:trimers}  
as a function of the interaction strength, confirming the existence of trimers at the FB point. While
in the strong coupling regime the pair binding energy can become arbitrary large, since $E_b\sim |U|$, this is not the case for trimers.   Due to the Pauli exclusion principle,  the pair and the extra fermion are separated by at least one lattice site, implying that the binding energy of the trimer must saturate to a constant value. Our DMRG calculations  indicate that 
$E_b^\textrm{trim}(U=-\infty)\approx 0.531$. 

\begin{figure}
	\includegraphics[width=\columnwidth]{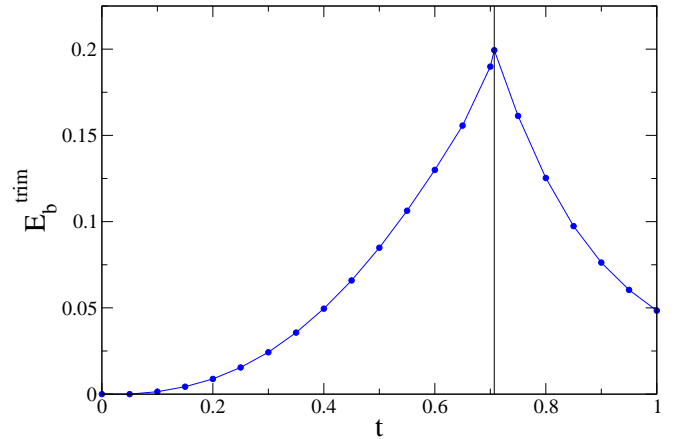}
	\caption{(Color online) Binding energy of the trimer as a function of the tunneling rate $t$ for $U=-10$.  The connecting line is a guide to the eye. The vertical line represents the flat band point $t=1/\sqrt{2}$.}
	\label{fig:3bodyEbvst}
\end{figure}

By continuity arguments, we expect that trimers exist for $t\neq 1/\sqrt 2$ provided $|U|$ is large enough. In Fig.~\ref{fig:3bodyEbvst}  we show the binding energy  of the trimer as a function of the tunneling rate $t$, for a fixed value of the interaction strength $U=-10$. 
	We see that $E_b^\textrm{trim}$  is peaked at the FB point, in complete analogy with the two-body binding energy. Moreover trimers break down near the integrable point, $t=0$, in agreement with the Bethe ansatz solution.

Trimers  in spin-1/2 fermionic systems arise from a subtle combination of  Hubbard interactions and tunneling processes, similarly to
 trions in semiconductors, where the constituent particles are the exciton (electron-hole pair) and an extra charge (electron or hole).	
When two spin up fermions are on neighboring sites, the spin down fermion can decrease its kinetik energy by delocalizing between them, without changing the double occupancy. In the linear chain geometry, this effect cannot occur, because the energy gain due to the delocalization of the spin-down fermion is exactly compensated by the energy cost to approach the two spin-up fermions. Trimers can nevertheless appear  
	if the spin-up component is  heavier than the spin-down counterpart, that is if it possesses a larger effective mass~\cite{Orso:PRL2010,Orso:CRAS2011,Roux:PRA2011,Dalmonte:PRA2012,Dhar:PRA2018} (see also ~\cite{Kartavtsev2009} for an equivalent result for continuous 1D models). A strong attractive atom-dimer interaction has indeed been  observed  experimentally \cite{Jag:PRL2014} in Fermi-Fermi mixtures of ultracold atoms with unequal masses, although in higher (three) dimensions.  
	Our DMRG results establish that the constraint of unequal masses  for the existance of trimers is no longer necessary in multiband systems.

We can better  understand the formation of trimers starting from the strong coupling regime $|U|\gg 1,t$.
 Since  
the pair is strongly bound, due to energy conservation it can never break in vacuum. We therefore consider a class of three-body states  
\begin{equation}\label{basis}
	e_r^{\sigma \sigma^\prime \dagger}	= \sum_i c_{i \uparrow}^{\sigma \dagger} c_{i\downarrow}^{\sigma\dagger} c_{i+r  \uparrow}^{\sigma^\prime\dagger} |\rangle, 
\end{equation}
describing a pair sitting in the $\sigma$ sublattice, with the extra spin-up fermion living in the sublattice  $\sigma^\prime$ at a distance $r$ from the dimer (the symbol $ |\rangle$ refers to the vacuum 
state). Notice that, due to the Pauli exclusion principle, we have $e_0^{AA\dagger}=e_0^{BB\dagger}=0$. 
We write the Hamiltonian (\ref{eq:Hbis}) as a sum of two terms, $H=H_t+H_U$, where  $H_t$ describes tunneling processes while $H_U$ accounts for the Hubbard interaction. We note that the  states in Eq.(\ref{basis})  are all  eigenstates of $H_U$ with eigenvalue $U$. The variational ground state energy of the trimer 
can then be obtained from  degenerate perturbation theory as
$E(2,1)\approx U+\lambda_t $, where $\lambda_t$ is  the lowest  eigenvalue of the following \emph{block} matrix
\begin{equation}
	\begingroup
	\mathcal{M}=
	\renewcommand*{\arraystretch}{1.5}
	\begin{pmatrix}\label{matrix}
		e_{i}^{AA}H_t e_r^{AA\dagger} & 	e_{i}^{AA}H_t e_r^{AB\dagger} & e_{i}^{AA}H_t e_r^{BA\dagger} & e_{i}^{AA}H_te_r^{BB\dagger}\\
		e_{i}^{AB}H_t e_r^{AA\dagger} & 	e_{i}^{AB}H_t e_r^{AB\dagger} & e_{i}^{AB}H_t e_r^{BA\dagger} & e_{i}^{AB}H_t e_r^{BB\dagger} \\
		e_{i}^{BA}H_t e_r^{AA\dagger} & 	e_{i}^{BA}H_t e_r^{AB\dagger} & e_{i}^{BA}H_t e_r^{BA\dagger} & e_{i}^{BA}H_te_r^{BB\dagger} \\
		e_{i}^{BB}H_t e_r^{AA\dagger} & 	e_{i}^{BB}H_t e_r^{AB\dagger} & e_{i}^{BB}H_t e_r^{BA\dagger} & e_{i}^{BB}H_t e_r^{BB\dagger} 
	\end{pmatrix}.
	\endgroup
\end{equation}

In order to evaluate $	\mathcal{M}$,  we need to know how the tunneling Hamiltonian acts on a generic state (\ref{basis}).  When  $H_t$ acts on the spin-up fermion forming the pair, it produces a state of zero double occupancy, which is orthogonal to the basis, so these processes can be neglected. The situation is different  when  $H_t$ acts on the spin-down fermion, because the latter could land on a neighboring site that is already occupied by the second spin-up fermion. In this case the pair and the extra fermion simply exchange 
their positions.
Within the subspace of one double occupancy, we find
\begin{eqnarray}
	H_t e_r^{AA\dagger}&=&	e_r^{AB\dagger} + e_{r+1}^{AB\dagger} \nonumber\\
	H_t e_r^{AB\dagger} &=& e_{r-1}^{AA\dagger}+  e_{r}^{AA\dagger}+t e_{r-1}^{AB\dagger}+t e_{r+1}^{AB\dagger}  \nonumber \\
	&-& \delta_{r,1}e_{-1}^{BA\dagger} - \delta_{r,0}e_0^{BA\dagger} \label{ansatz}\\
	H_t e_r^{BA\dagger} &=&- \delta_{r,0} e_0^{AB\dagger}-\delta_{r,-1}e_1^{AB\dagger}+ e_r^{BB\dagger}+ e_{r+1}^{BB\dagger}\nonumber\\
	H_t e_r^{BB\dagger} &=&e_{r-1}^{BA\dagger} +e_{r}^{BA\dagger} +t e_{r-1}^{BB\dagger}+t e_{r+1}^{BB\dagger}\nonumber \\ &-&t\delta_{r,1} e_{-1}^{BB\dagger}-t\delta_{r,-1} e_1^{BB\dagger}.\nonumber
\end{eqnarray} 
The terms in the rhs of Eq.~(\ref{ansatz}) with  negative sign originate from the exchange processes between the pair and the extra spin-up fermion, which can only occur if $r=0$ or $r=\pm 1$. For instance the first negative term in the third line of Eq.~(\ref{ansatz}) comes from $\sum_j c_{j+1\downarrow}^{B\dagger} c_{j\downarrow}^{A} e_r^{AB\dagger}=\sum_i c_{i\downarrow}^{A\dagger} c_{i+1\downarrow}^{B\dagger}c_{i+r \uparrow}^{B\dagger} |\rangle=-\delta_{r,1}e_{-1}^{BA\dagger}$.

\begin{figure}
	\includegraphics[width=\columnwidth]{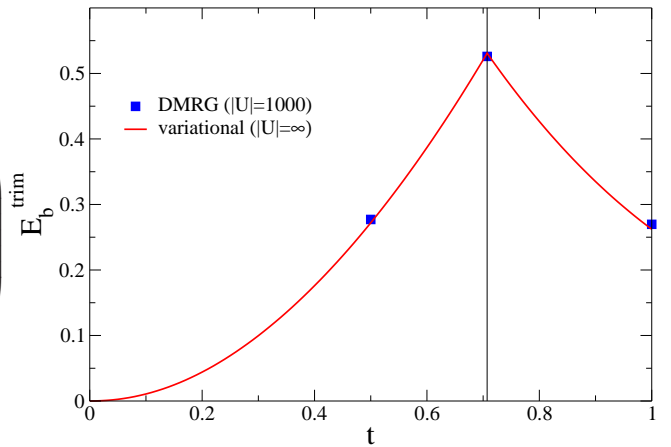}
	\caption{(Color online) Binding energy of the trimer as a function of the tunneling rate $t$. The solid line is the
		prediction of the variational approach for infinite attraction, obtained by diagonalizing the Hamiltonian over the 
		class of three-body states in Eq.~(\ref{basis}). The square symbols are DMRG results  for three different values of $t=1/2, 1/\sqrt 2, 1$ and  $U=-1000$.  
		The vertical line represents the flat band point $t=1/\sqrt{2}$.}
	\label{fig:3bodyEbvst_var}
\end{figure}

By using  Eq.~(\ref{ansatz}) it is now straightforward to evaluate all the entries of $	\mathcal{M}$. Notice that many blocks are actually null matrices; for instance in the first row of Eq.~(\ref{matrix})  only one block (the second)  is  nonzero. We
diagonalize the matrix (\ref{matrix}) numerically after introducing a cut-off integer $N$ for the relative distance between the pair and the extra spin-up fermion, thus limiting the size of each block according to $|i|, |r|\leq N$.
For given $N$,  $\mathcal{M}$ is a square matrix of dimension $8N+2$. We extract its lowest eigenvalue by choosing $N$ large enough to ensure full convergence and extract the binding energy from Eq.~(\ref{def:Ebtrim}).
The obtained results as a function of the tunneling rate are shown in Fig.~\ref{fig:3bodyEbvst_var} by the solid line.  
We see that our variational approach for $U=-\infty$ reproduces all the expected features, 
notably the absence of trimers for $t=0$ and the peak in the  three-body binding energy at the FB point.  
For $t=1/\sqrt 2$ 
it  gives $E_b^\textrm{trim}\approx 0.53087$, which is in very good agreement with the DMRG data for $U=-1000$ (square symbols). Moving away from the FB point, the variational approach slightly underestimates the  binding energy of the trimer. A fit to the numerical results for small $t$ reveals that the binding energy of the trimer, calculated within the variational approach, vanishes as $t^2$ approaching the integrable point, $t=0$.  This result implies that, for infinite attraction, trimers exist for \emph{any} nonzero value of $t$.

Our numerics shows that the convergence of the binding energy as a function of the cut-off $N$ is particularly fast approaching the FB point. Indeed  using $N=1$ yields $E_b^\textrm{trim}\approx 0.51317$, corresponding to a relative error  of only $3\%$. In this case the matrix   in Eq.~(\ref{matrix})  reduces to the $10\times 10$ matrix
\begin{equation}
	\mathcal{M}=
	\left(
	\begin{array}{cccccccccc}
		0 & 0 & 1 & 1 & 0 & 0 & 0 & 0 & 0 & 0 \\
		0 & 0 & 0 & 0 & 1 & 0 & 0 & 0 & 0 & 0 \\
		1 & 0 & 0 & t & 0 & 0 & 0 & 0 & 0 & 0 \\
		1 & 0 & t & 0 & t & 0 & -1 & 0 & 0 & 0 \\
		0 & 1 & 0 & t & 0 & -1 & 0 & 0 & 0 & 0 \\
		0 & 0 & 0 & 0 & -1 & 0 & 0 & 0 & 1 & 0 \\
		0 & 0 & 0 & -1 & 0 & 0 & 0 & 0 & 0 & 1 \\
		0 & 0 & 0 & 0 & 0 & 0 & 0 & 0 & 0 & 1 \\
		0 & 0 & 0 & 0 & 0 & 1 & 0 & 0 & 0 & -t \\
		0 & 0 & 0 & 0 & 0 & 0 & 1 & 1 & -t & 0 \\
	\end{array}
	\right).
\end{equation}
By moving away from the FB point, the value of  $N$ needed to ensure convergence becomes larger and larger, signaling that the mean distance between the pair and the extra fermion increases and the binding energy is reduced.
  
\section{Conclusion and outlook}
\label{sec:conclusions}

In this work we have investigated the Fermi Hubbard model with attractive interactions on the 1D sawtooth lattice. From the solution of the two-body problem, we have extracted the binding energy and the effective mass of the pair, both analytically and numerically. We have shown that,
in a broad region of $t$ values around the FB point, both quantities are highly sensitive to weak interactions.  In particular, the binding energy possesses a pronounced maximum in correspondence of the FB point, $t=1/\sqrt 2$, which persists for any  $U<0$. From the inverse effective mass of the pair at the FB point we have estimated the superfluid weight $D_s$ of the many-body system, showing that it is 
in good agreement with the DMRG calculations of \cite{chan2021pairing}.

Our numerical results for fully-paired many-body systems reveal that the proximity to a flat band significantly modifiy the nature of the BCS-BEC crossover.
 While  the chemical potential remains always pinned near its two-body value,  the  pairing gap is strongly depleted at finite density  and
takes its maximum value not at the FB point, but at a shifted position $t^*<1/\sqrt 2$, which depends on the value of the density. 
	We show that the anomalous pairing in the sawtooth lattice comes from the fact that the energy change upon adding an extra spin-up fermion to the system falls
	 below the bottom of the single-particle spectrum, 
	$\mu_\uparrow < E_\textrm{ref}/2$, causing the appearance of an effective attraction between the pairs in the medium and the excess fermion. Importantly, we have unveiled that  two spin-up and one spin-down fermions in the sawtooth lattice  can form a three-body bound state, whose binding energy is also peaked at the FB point and vanishes at the integrable point, $t=0$. 
	Our results establish that trimers exist in flat band lattices and they are detrimental to superconductivity.

It would be interesting to study by exact numerics the behavior of the superfluid  weight $D_s$ in the sawtooth lattice for a generic tunneling rate and its relation with the pair inverse effective mass. Our results show that  for finite $|U|$ the minimum of $1/m_p^*$ drifts towards  smaller values of $t$. 
Another intriguing direction is to understand whether multiband BCS theory can correctly predict the anomalous behavior of the pairing gap observed in our numerics, especially at low density. 

The results discussed in this work can be investigated experimentally with cold atoms in optical lattices. In particular a viable scheme to implement the sawtooth lattice  has been recently proposed
\cite{Huber:PRB2010,Zhang2015}. 
The interaction strength  can be controlled either directly, via a Feshbach resonance or indirectly, by
varying the  tunneling rates $t, t^\prime$ and consequently the ratios  $t/U$ and $t^\prime/U$.
While we have mainly focused on the sawtooth lattice, we expect that our results will apply also to other FB systems.

\emph{Note added: } The existence of trimers in the 1D sawtooth lattice at the FB point has been
confirmed in a very recent preprint~\cite{Iskin:arXiv2022} by Iskin, reporting a very good agreement with our
DMRG results displayed in Fig.~\ref{fig:trimers}. In the preprint the three-body problem is solved numerically by mapping it into an effective integral equation~\cite{Mattis:RMP1986,Orso:PRL2010} and  the  presence of trimers for other values of the tunneling rates  has also been discussed.

\section*{ACKNOWLEDGEMENTS} 
We thank S. Pilati and F. Chevy for useful comments on the manuscript.
G.O. acknowledges financial
support from ANR (Grant SpiFBox) and from DIM
Sirteq (Grant EML 19002465 1DFG).
M.S. acknowledges funding from MULTIPLY fellowship under the Marie Sk\l{}odowska-Curie COFUND Action (grant agreement No. 713694).

\input{ms_rev.bbl}	

\end{document}

%% file: ms_rev.bbl
%